# Geometric symmetry breaking and nonlinearity can increase thermoelectric power

Jonatan Fast,[1,2] Hanna Lundström,[1,2] Sven Dorsch,[1,2] Lars Samuelson,[1,2,4] Adam Burke,[1,2] Peter Samuelsson,[1,3] Heiner Linke[1,2]

[1] NanoLund, Lund University, Box 118, 22100 Lund, Sweden
[2] Solid State Physics, Lund University, Box 118, 22100 Lund, Sweden
[3] Mathematical Physics, Lund University, Box 118, 22100 Lund, Sweden
[4] Institute of Nanoscience and Applications, Southern University of Science and Technology, Shenzhen, China

**Direct thermal-to-electric energy converters typically operate in the linear regime, where the ratio of actual maximum power relative to the ideal maximum power, the so-called fill factor (*FF*), is 0.25. By increasing the *FF* one can potentially increase maximum power by up to four times, but this is only possible in the nonlinear regime of transport and has previously rarely been considered. Here we show, based on fundamental symmetry considerations, that the leading order non-linear terms that can increase the *FF* require devices with broken spatial symmetry. To experimentally demonstrate such a system, we study nonlinear, thermoelectric transport across an asymmetric energy barrier epitaxially defined in a single semiconductor nanowire. We find in both experiment and theory that we can increase the *FF* above the linear-response limit of 0.25, accompanied by a drastic increase in short circuit current, open-circuit voltage and maximum power. Our results show that geometric symmetry breaking combined with the design of nonlinear behaviour represent a design strategy for increasing the performance of thermal-to-electric energy converters such as in hot-carrier photovoltaics, thermophotovoltaics or in anisotropic thermoelectric materials.**

The ability to effectively convert heat stored in charge carriers into electricity is at the heart of existing and emerging technologies ranging from thermoelectric generators to hot-carrier photovoltaics and thermophotovoltaics. Generally, such devices operate in the linear-response regime where power output is fundamentally limited [1–3]: the so-called fill factor *FF,* which describes the shape of the current-voltage (*I-V*) curve in the power-producing quadrant, has the value *FF* = 0.25 by definition in the linear response regime. This means that the device's maximum power is four times smaller than the maximum power that corresponds to an ideal *FF* = 1 [4]. Optimizing the *FF* is a well-established strategy to increase the power output in photovoltaics, where a good silicon single-junction cell typically reaches *FF* ≈ 0.8 [5]. This strategy is, however, rarely considered in the context of thermal-to-electric energy conversion.

To achieve *FF* > 0.25 in thermal-to-electric energy conversion, it is necessary to introduce nonlinear *I-V* behaviour. What nonlinear features are the most promising for increasing the *FF*? In the present work we show, based on fundamental symmetry considerations, that the leading order nonlinear terms that can increase the *FF* require devices that respond asymmetrically to the direction of external bias. This observation highlights device (a-)symmetry as a critical attribute for increasing thermoelectric power.

To experimentally demonstrate the role of broken symmetry, we study thermoelectric transport, in the nonlinear regime, across an asymmetric, ramp-shaped, energy barrier. The barrier is epitaxially defined by heterostructure engineering in a single semiconductor nanowire (Fig. 1a-c). The nanowire is equipped at either end with independent heaters that allow us to control the thermal bias in both directions along its axis (Fig. 1d,e). We observe a large asymmetry in thermoelectric response with respect to the direction of thermal bias. Crucially, in one of the two configurations, we observe a *FF* > 0.25 that increases linearly in thermal bias, qualitatively consistent with theoretical predictions based on fundamental symmetry considerations. The strategy introduced here, to design device symmetry and its nonlinear behaviour in order to increase the *FF*, offers new avenues for increasing the performance of thermal-to-electric energy converters.

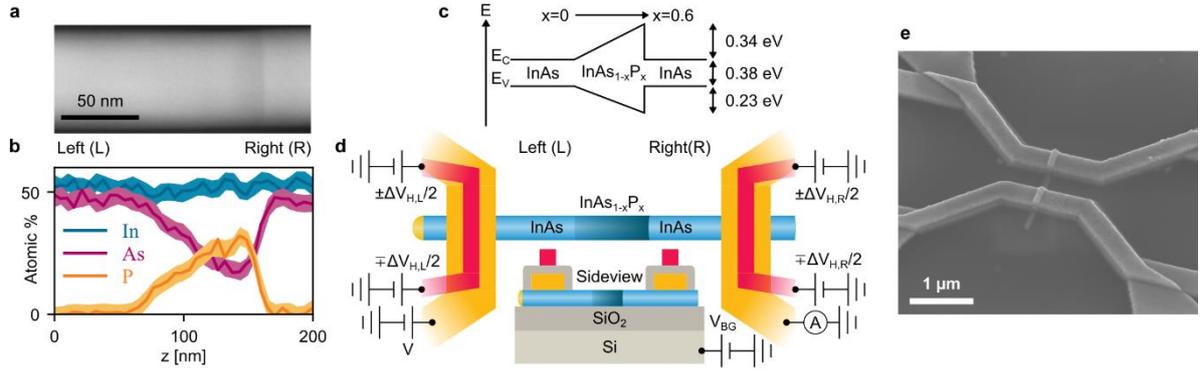

**Figure 1. a**, Scanning transmission electron microscope image and **b**, energy-dispersive X-ray spectroscopy of a typical nanowire from the same growth as the studied nanowire, indicating the energy barrier shape. **c**, Approximate, idealised, structure of the conduction (valence) band, $E_C$ ($E_V$). **d**, Device layout. A bias voltage, $V$, can be applied along the nanowire using the metallic leads (yellow). A heating voltage $\Delta V_{H,L}$ ($\Delta V_{H,R}$) can be applied on the left (right) side independently via the electrically insulated top heaters (red). The side view shows oxide layers for insulation on top of p-type SI for application of a global back-gate voltage ($V_{BG}$). **e**, Scanning electron micrograph of the completed device.

### Symmetry properties, nonlinear response and fill factor

We begin by considering fundamental symmetry properties of a generic, n-type, two-terminal thermoelectric device driven by applying a thermal bias $\Delta T_{(L/R)}$ to the left (L) or the right (R) contact, respectively, resulting in an electrical current $I(V, \Delta T_L, \Delta T_R)$. For heating at the right contact only ($\Delta T_R = \Delta T > 0$, $\Delta T_L = 0$) the current can be expanded to lowest nonlinear order in voltage and thermal bias as

$$I(V, 0, \Delta T) = GV + L\Delta T + MV^2 + N\Delta T^2 + H_R V \Delta T, \quad (1)$$

where $G$, $L$, $M$, $N$, and $H$ are constants describing the particular system. Heating instead at the left contact, assuming that heating by itself does not alter the device properties, that is $I(0, 0, \Delta T) = -I(0, \Delta T, 0)$, we can expand the current

$$I(V, \Delta T, 0) = GV - L\Delta T + MV^2 - N\Delta T^2 + H_L V \Delta T. \quad (2)$$

For a spatially symmetric device it further holds that the current reverses sign under a simultaneous reverse of bias $V$ and swapping of heating contact, that is

$$I(V, 0, \Delta T) = -I(-V, \Delta T, 0). \quad (3)$$

A symmetric device, fulfilling equation (3), constrains the expansion coefficients in equations (1) and (2) to $M = 0$ and $H_L = H_R$.

The fill factor is defined as $FF = P_{max}/I_{SC}V_{OC}$, where $P_{max}$ is the electrical power maximized with respect to $V$, $I_{SC}$ the short-circuit current, and $V_{OC}$ the open-circuit voltage. From the nonlinear current expressions in equation (1) and (2) we evaluate $P_{max}$, $I_{SC}$, and $V_{OC}$ (see Methods). The resulting fill factor, expanded to lowest nonlinear order in thermal bias, is

$$FF^{(L/R)} = \frac{1}{4} \mp \frac{ML}{8G^2} \Delta T, \quad (4)$$

with +/- corresponding to heating at L/R, respectively. Hence, we can make our first key observation: for a spatially symmetric device ($M = 0$), to lowest order nonlinearity (equation (1) and (2)), the fill factor is restricted to its linear response value, $FF = 0.25$. Increasing the $FF$ of a thermoelectric system using leading order nonlinear behaviour thus requires a system with broken spatial symmetry, such that $M \neq 0$, suggesting symmetry-breaking as a novel strategy for increasing the $FF$ and thus thermoelectric power.

**Observing asymmetric thermoelectric transport**

To our knowledge, an asymmetric thermoelectric system has not previously been experimentally demonstrated. It is however well established that nonlinear thermoelectric systems can be realised in mesoscale devices [6,7], potentially with asymmetric behaviour such as thermal rectification [8–11]. One approach to realise such systems is by varying the chemical composition of semiconductor materials at the nanometer scale in order to create potential barriers with energy selective transmission [12–15]. In semiconducting nanowires, such heterostructures can be synthesised with high control and precision [16,17]. Indeed, highly non-linear thermoelectric devices have previously been realised in single nanowire systems [18]. In this work, a geometrically asymmetric potential barrier is epitaxially defined as a heterostructure in a nanowire.

Our experimental device consists of a single InAs nanowire with an $InAs_{1-x}P_x$ segment where the ratio of P to As is gradually increased to yield a ramp-shaped energy barrier, both in conduction and valence band, before abruptly transitioning back to InAs (Fig. 1c). We expect the carrier transport will be dominated by the behaviour of electrons in the conduction band, as the chemical potential at InAs surfaces are known to be pinned in the conduction band [19,20]. The presence and shape of the heterostructure is confirmed by energy-dispersive X-ray spectroscopy (Fig. 1b). Both ends of the nanowire are contacted by electrical leads, on top of which we place electrically insulated top heaters [21,22] (see Fig. 1d,e). Applying a heating voltage $\Delta V_H$ to one of the top heaters generates Joule heat. The heat is transferred to the corresponding side of the nanowire via thermal conduction through the electric leads, resulting in a local increase of temperature $\Delta T$. Using this design, thermal and electrical biasing can be applied independently from one another, and selectively in either direction across the barrier.

We define the left (L) and right (R) terminals of the device as the sides where the barrier has gradient-like shape, and a steep side, respectively (Fig. 1c,d). Electrical bias is applied on the left side and current measured on the right, such that $I$ is positive when electrons flow from R to L (Fig. 1d). For thermal-bias measurements, $\Delta V_H$ is applied on either the left side ($\Delta V_{H,L}$) or the right side ($\Delta V_{H,R}$), resulting in a corresponding local temperature increase in the left ($\Delta T_{H,L}$) or right ($\Delta T_{H,R}$) end of the nanowire. The experiments were performed at a base temperature of $T_0$ = 77 K, meaning that the total temperature is given as $T_{(R/L)} = T_0 + \Delta T_{(R/L)}$.

We find that *I-V* curves in the absence of thermal bias ($\Delta V_{H,(L/R)}$ = 0) are nonlinear already at $V$ of a few mV (Fig. 2a). Increase of $\Delta V_{H,L}$ ($\Delta V_{H,R}$) generates an additional negative (positive) thermal current, resulting in a downward (upward) shift of the *I-V* curve. The sign of the thermal current indicates that heating on one side of the barrier results in a net electron flow to the opposite side. As $\Delta V_{H,L}$ ($\Delta V_{H,R}$) is increased, the curve shape becomes more concave (convex) in the power-producing quadrant (the region between $I_{SC}$ and $V_{OC}$). The change in curve shape suggests an increase in *FF* for heating on the right side, and a decrease in *FF* for heating on the left side.

To validate our results, and to determine the relation of $\Delta V_H$ to $\Delta T$, we model the experiment within a Landauer-Büttiker scattering framework [23], fully incorporating nonlinear bias effects [24–26] (see methods). Adaptations to the experimental data are found with strong agreement for all *I-V* curves (black solid lines in Fig. 2a), using a single set of parameters (see caption to Fig. 2a). The agreement of the adaptations provides evidence that the experimentally observed behaviour is indeed due to the asymmetric barrier, and not, for example, due to imperfections in the nanowire. Furthermore, we can determine values for $\Delta T$ for each $\Delta V_H$ (inset to Fig. 2a). The adaptation indicates an equilibrium chemical potential $\mu_0$ on the order of 100 meV. Given the small experimental thermal energy of $k_B T_0 \approx 7$ meV, electron energies $E$ in all experiments reported here are thus expected to be well below the barrier top ($E < U_{top}$), such that the device operates in the tunnelling-transport regime.

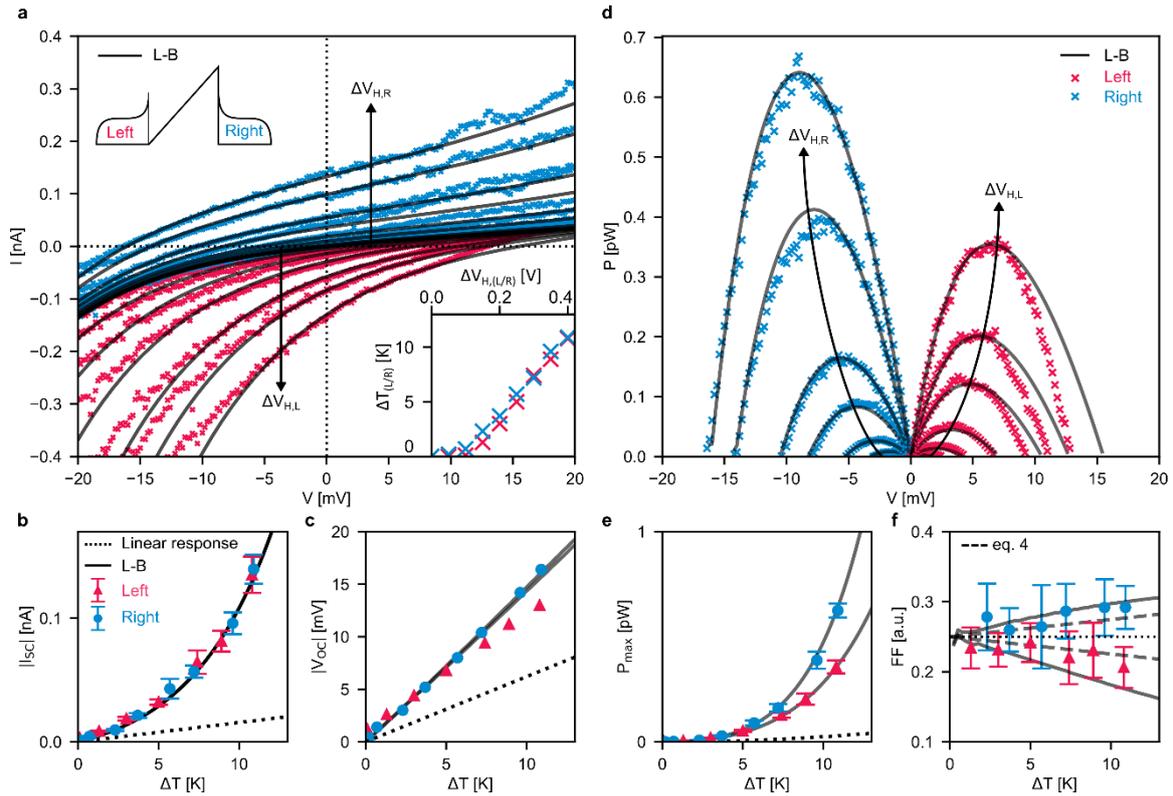

**Figure 2. a,** *I-V* curves while stepwise increasing the heating voltages $\Delta V_{H,R}$ (blue marks), or $\Delta V_{H,L}$ (red marks), respectively. Arrows indicate the direction curves shift for increasing $\Delta V_H$. Each data point is the result of averaging 6 sweeps (standard deviations are shown in Supplementary Fig. 3). Solid black lines labelled L-B are calculated using Landauer-Büttiker theory (equation (M12)) for barrier height $U_{top}$ = 340 meV, barrier length $l$ = 92 nm, base temperature $T_0$ = 77 K, equilibrium chemical potential $\mu_0$ = 100 meV, and scalling factor $A$ = 710. Inset shows relationship between $\Delta V_H$ and the resulting temperature bias $\Delta T$, where values for $\Delta T$ are extracted from fits of equation (M12) to *I-V* curves in (**a**). **b,c,** Short-circuit current $I_{SC}$ and open-circuit voltage $V_{OC}$ as a function of $\Delta T$. Error bars indicate standard deviation, originating from the averaging of 6 sweeps (Supplementary Fig. 3), which in (**c**) are smaller than the data point markers. **d,** Power $P = |IV|$ in the power-producing quadrants as a function of $V$. **e,f,** Maximum output power $P_{max}$ and fill factor $FF$ as a function of $\Delta T$. The experimental values for $P_{max}$ are extracted from second order polynomial fits to curves in (**d**), see Supplementary Fig. 3b. The dashed line in (**f**) shows $FF$ calculated from equation (4) using the values for $G$, $L$, and $M$ extracted from experimental data (Supplementary table 1). Dotted lines in (**b**, **c**, **e**, and **f**) show expected values under linear response, based on the extracted values for $G$ and $L$.

**Spatial asymmetry and nonlinearity can increase the fill factor**

Importantly, the relationship between $\Delta V_H$ and $\Delta T$ indicates that the experimental heating arrangement operates symmetrically: we observe the same $\Delta T$ when $\Delta V_{H,L} = \Delta V_H$ and $\Delta V_{H,R} = 0$, as when $\Delta V_{H,L} = 0$ and $\Delta V_{H,R} = \Delta V_H$. Any observed asymmetries in the *I-V* curves are thus not expected to be related to unintentional asymmetries in the heater arrangement or performance. This observation is consistent with the symmetric behaviour of $I_{SC}$ (Fig. 2b). At short-circuit conditions, and for symmetric heating, the response is expected to be symmetric with respect to heating side irrespective of the direction of thermal bias, since the transmission probability across the barrier should be unchanged by temperature.

One main purpose of our study is to test whether device asymmetry and nonlinear behaviour can be used to increase the $FF$ beyond the linear response limit. To determine the experimental $FF$, we compared the ratio between the product $I_{SC}V_{OC}$ (Fig. 2b, c)) to $P_{max}$, the maximum value of $P = |IV|$, for each $\Delta T$ (Fig. 2d,e). Most interestingly, when heating terminal R we find a quasi-linear fill-factor increase up to $FF$ = 0.29 ± 0.03 at $\Delta T \approx 10$ K, an almost 20% improvement compared to the linear-response limit *FF* = 0.25 (Fig. 2f). When heating terminal L, we observe instead a corresponding decrease to *FF* = 0.21 ± 0.03. The observed quasi-linear behaviour of the *FF* with Δ*T*, as well as the observed splitting of *FF* with respect to heating direction, is in good qualitative agreement with the fundamental, lowest-order nonlinear prediction for a system where $M \neq 0$ (equation (4)).

To check for quantitative agreement with equation (4), we extract values for all coefficients in the generic nonlinear expansions of equation (1) and (2) (Supplementary Table 1). Details of this process are described in supplementary section IV. Briefly, *G* and *M* are extracted from *I-V* curves measured at Δ*T* = 0 (Fig. 3a,b), *L* and *N* are extracted from *I*(Δ*T*) measured at *V* = 0 (Fig. 3c,d), and *H* is determined from measurements at finite *V* and Δ*T* in the linear response limit (Fig. 3e,f). The origin of the corresponding nonlinear effects is illustrated in Fig. 3.

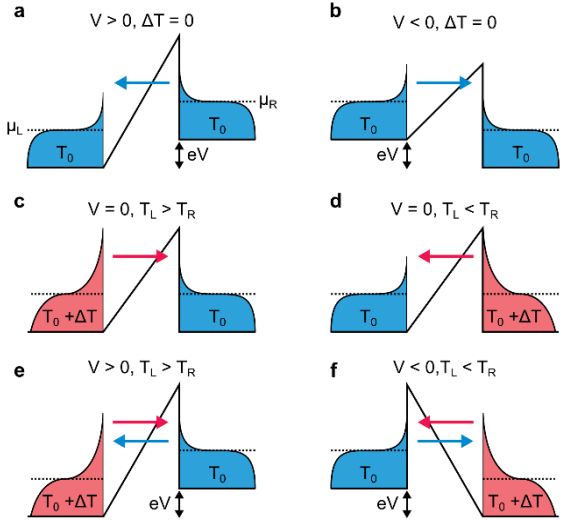

**Figure 3.** Illustration of possible causes of the nonlinear effects and associated asymmetric behaviour observed here. Blue and red coloured regions illustrate Fermi-Dirac distributions. For consistent sign convention, it is assumed that the electrical bias is applied on the left side (as indicated in Fig. 1d). **a,b**, Nonlinear effects originating from electrical bias during tunnelling transport. An electrical bias *V* shifts *μ* by e*V* and causes electrons to flow to the side with lower *μ*, but alters also the barrier shape, such that $I(V) \neq -I(-V)$ for finite *V*. To lowest nonlinear order, these effects are described by $I = GV + MV^2$. **c,d**, Nonlinear effects originating from temperature-induced changes in the energy range of carriers participating in transport, described by the Fermi-Dirac distributions. Increasing temperature on one side causes electrons to flow to the colder side because electron transmission is higher at higher energy. To lowest nonlinear order, these effects are described by $I = L\Delta T + N\Delta T^2$. **e,f**, When producing power, the device is operated under both thermal and electrical bias, such that nonlinear transport of both origins may occur, and all five terms in equation (1,2) are needed for the lowest nonlinear order description.

Using the values for *M, L,* and *G* from Supplementary Table 1, equation (4) somewhat underestimates the observed *FF* splitting (dashed lines, Fig. 2f). Comparison between equation (4) and Landauer-Büttiker model, which includes full non-linear behaviour, indicates that about half of the experimentally observed *FF* change is accounted for by the lowest-order nonlinear prediction (equations (1) and (2)), whereas higher-order nonlinear terms account for the rest.

In addition to the change in *FF*, which describes a relative power, we also observe a drastic increase in absolute power as a function of Δ*T* (Fig. 2e). The observed $P_{max}$ in the nonlinear regime at Δ*T* = 10 K ($\Delta T/T_0 \approx 13\%$) is up to 20 times higher than the $P_{max,lin} = (L^2/4G)\Delta T^2$ predicted by the device's linear response properties (Fig. 2e). The major part of this increase can be attributed to the clearly nonlinear behaviour of $I_{SC}$ (Fig. 2b), which in the observed range of Δ*T* is well described by $I_{SC} = L\Delta T + N\Delta T^2$ (Supplementary Fig. 10c). A smaller portion of the increase can be attributed to an increase in $V_{OC}$, which also exceeds the linear response prediction $V_{OC} = (L/G)\Delta T$ (Fig. 2c). At Δ*T* = 10 K, we find

$V_{OC}/\Delta T \approx 1.5$ mV/K, large compared to commonly observed Seebeck coefficients in linear response both in bulk and nanowire InAs [27–29].

## 6. Conclusion and outlook

We have shown that fundamental symmetry relationships dictate that samples with broken geometrical symmetry are a pre-requisite for observing leading-order nonlinear effects that could increase the thermoelectric *FF*. We also experimentally demonstrated the leading-order asymmetric change in *FF* predicted by equation (4) for an asymmetric device. This observation explains why nonlinear *I-V* curves are rarely observed in conventional thermoelectric systems, which typically are isotropic.

Although transport across our ramp-shaped energy barrier was in the tunnelling regime, where a very low current is expected, we nevertheless observe thermoelectric power on the order of 1 pW at $\Delta T \approx 10$ K, comparable to that observed in plain InAs nanowires of comparable quality and diameter at similar temperatures [30]. Future studies using different barrier shapes and varied $\mu_0$ may be used to explore whether power production and *FF* can be further increased.

Importantly, our results suggest that improving the thermoelectric *FF* is a promising approach for creating more powerful thermoelectric devices and materials. Looking to photovoltaic devices for inspiration, where a *FF* > 0.75 is not unusual, we can conclude that, for a given device or material, there is potential for an up to three-fold increase in maximum power just based on *FF*, and more if nonlinear increases in $I_{SC}$ and $V_{OC}$ can be realised. In order to leverage the required nonlinear effects to leading order, devices with broken spatial symmetry must be used.

There exists a wealth of systems that may be investigated to explore this design route for use in direct thermal-to-electric energy conversion. Examples include mesoscopic devices other than the one used in this study, including layered, two dimensional systems commonly used in hot-carrier photovoltaics [31–33]. A particularly promising candidate might be monolayers of asymmetric molecules [34,35]. Another route might be the design of anisotropic materials or metamaterials that combine asymmetry with nonlinear properties.

**Acknowledgement**
We thank M. Kumar for the synthesis of the nanowires, D. Madsen for the nanowire imaging with transmission electron microscope and compositional analysis with energy dispersive X-ray spectroscopy, and R. S. Whitney for valuable discussions. This work was supported by the Knut and Alice Wallenberg Foundation (Project No. 2016-0089), by the Swedish Research Council (grant number 2018-03921), and by NanoLund.

**Author contributions**
H. Li. conceived the study and guided it jointly with P.S.. L.S. designed the 1D potential structures and supervised nanowire growth. H. Lu., J.F., and A.B. designed and fabricated the devices and carried out the experiment. J.F., H. Lu., S.D., P.S., H. Li., and A.B. analysed and interpreted the data. P.S. and J.F. performed theoretical calculations and adaptations. All authors contributed to writing and editing the manuscript.

**Method**
*Short-circuit current, open-circuit voltage and maximum power.*
Based on equation (1) and (2), equation (4) is derived from the expressions for $I_{SC}$, $V_{OC}$, and $P_{max}$. For heating $\Delta T$ at the right contact (R) the nonlinear expansion of the current $I(V, 0, \Delta T)$ is given by equation (1) in the main text. The short-circuit current, defined as the current at $V = 0$, is then

$$I_{SC}^{(R)} \equiv I(0,0,\Delta T) = L\Delta T + N\Delta T^2. \qquad (M1)$$

The corresponding open-circuit voltage, obtained from $I(V_{OC}^{(R)}, 0, \Delta T) = 0$, is

$$V_{OC}^{(R)} = \frac{\sqrt{(G + H_R \Delta T)^2 - 4M\Delta T(L + N\Delta T)} - (G + H_R \Delta T)}{2M}. \quad (M2)$$

The electrical power $I(V, 0, \Delta T)V$ is maximized for a voltage $V_{max}^{(R)}$ given by

$$V_{max}^{(R)} = \frac{\sqrt{(G + H_R \Delta T)^2 - 3M\Delta T(L + N\Delta T)} - (G + H_R \Delta T)}{3M}. \quad (M3)$$

The corresponding maximal power is

$$P_{max}^{(R)} = |I(V_{max}^{(R)}, 0, \Delta T) V_{max}^{(R)}|. \quad (M4)$$

For heating $\Delta T$ at the left contact (L) quantities corresponding to equation (M1) to (M4) become

$$I_{SC}^{(L)} \equiv I(0, \Delta T, 0) = -L\Delta T - N\Delta T^2, \quad (M5)$$

$$V_{OC}^{(L)} = \frac{\sqrt{(G + H_L \Delta T)^2 + 4M\Delta T(L + N\Delta T)} - (G + H_L \Delta T)}{2M}, \quad (M6)$$

$$V_{max}^{(L)} = \frac{\sqrt{(G + H_L \Delta T)^2 + 3M\Delta T(L + N\Delta T)} - (G + H_L \Delta T)}{3M}, \quad (M7)$$

$$P_{max}^{(L)} = |I(V_{max}^{(L)}, \Delta T, 0) V_{max}^{(L)}|. \quad (M8)$$

In the limit of weak nonlinearity, we can expand the open circuit voltage and the maximum power to first nonlinear order in $\Delta T$. This gives

$$V_{OC}^{(L/R)} = \pm \frac{L}{G} \Delta T - \left( \frac{ML^2 \mp (G^2 N - GLH_{L/R})}{G^3} \right) \Delta T^2, \quad (M9)$$

$$P_{max}^{(L/R)} = \frac{L^2}{4G} \Delta T^2 \mp \frac{L}{8G^3} \left( ML^2 \mp \left( 4G^2 N - 2GLH_{\frac{L}{R}} \right) \right) \Delta T^3, \quad (M10)$$

where upper/lower sign in $\pm, \mp$ corresponds to L/R. Together with $I_{SC}$ in equations (1) and (5), we can calculate the fill factor to first nonlinear order as

$$FF^{(L/R)} = \frac{P_{max}^{(L/R)}}{I_{SC}^{(L/R)} V_{OC}^{(L/R)}} = \frac{1}{4} \mp \frac{ML}{8G^2} \Delta T, \quad (M11)$$

which is the expression in equation (3) in the main text.

*Nanowire growth*

The nanowires were grown in a custom made Chemical Beam Epitaxy (CBE) system operating under ultra-high vacuum conditions, with the use of trimethylindium (TMI), tertiarybutylarsine (TBAs) and tertiarybutylphosphine (TBP) as the sources of In, As and P, respectively. The TBAs and TBP gas sources were passed through a cracker just prior to injection into the CBE growth chamber. Aerosol-created catalytic gold particles with a diameter of approximately 50 nm are used as growth-seeds, resulting in nanowire diameters of 65 (±6) nm (determined by transmission electron microscope). The ramp-like heterostructure was achieved in a step-wise manner by repeatedly pre-selecting a designed As-to-P ratio prior to initiating the In-flow to grow the next "step", as pioneered in earlier work by Nylund et al. [36]. Gradually increasing P-to-As ratios, starting from 0, a fine stair-case was realized, effectively operating as a quasi-continuous graded ramp. At the highest point of the ramp, an abrupt transition to the binary InAs is performed, again after setting up the group-V flow to As-only, before re-initiating the growth by starting the flow of Indium.

The composition in atomic percentage along the nanowire axis is determined via energy-dispersive X-ray spectroscopy (EDX). By averaging over eight different nanowires from the same growth (Supplementary Fig. 1), it is determined that the atomic composition gradually changes over a distance of $L$ = 92 (±7) nm from InAs towards InP along the nanowire axis ($z$-direction). The composition at the peak is estimated to InAs$_{0.4}$P$_{0.6}$, before abruptly transitioning to pure InAs within approximately 10 nm. Based on the known band offsets between InAs and InP, [37] and assuming that the offsets changes linearly with x in InAs$_{1-x}$P$_x$, [38] the barrier height, $U_{top}$, is estimated to 340 (±4) meV.

*Device/fabrication*

Nanowires are mechanically deposited from the growth chip to lie horizontally on a p++ doped Si substrate covered by a 117 nm thick SiO2 layer, allowing the Si to serve as a global back gate. Metallic leads are fabricated to each end of the nanowire via electron beam lithography (EBL, Raith-150), sulphur passivation, [39] and evaporation of a 25 nm Ni adhesion layer followed by 75 nm Au using a Temescal e-beam evaporator. An insulating HfO$_2$ -layer (80 atomic layers) is deposited via atomic layer deposition (ALD) on top of the device using an Oxford Instruments Savannah ALD system. A second cycle of EBL and metal evaporation is performed to fabricate the electrically insulated top heaters. For further details on the fabrication see ref. [21,40].

*Measurement*

For electrical characterization, the device was submerged in liquid nitrogen ($T_0$ = 77 K, $kT_0 \approx$ 6.7 meV) to reduce thermal noise. At this temperature, the conductance is relatively low (Supplementary Fig. 2), wherefore $V_{BG}$ = 10.3 V was applied for the measurements presented. For *I-V* measurements, *V* is applied using a Yokogawa 7651, and a Femto DLPCA-200 current pre-amplifier is used to measure *I* (Fig. 1d). The heating voltage, $\Delta V_H$, is applied symmetrically as indicated in Fig. 1d, applying +(-) $\Delta V_H$/2 on respective ends of the heater. This way of applying the voltage ensures that the potential at the point of the heater is roughly 0 V, so that the top heaters do not alter the potential landscape of the nanowire via unintentional gating effects.

*Landauer-Büttiker model*

We model the electrical current $I$ flowing from contact L to R of the device (Fig. 1c) within a Landauer-Büttiker scattering framework [23], fully incorporating non-linear bias effects [24]. The contact chemical potentials are $\mu_L = \mu_0 - eV$ and $\mu_R = \mu_0$, with $\mu_0$ the equilibrium value, $V$ the applied voltage bias, and $e$ the electron charge. The current $I(V, \Delta T_L, \Delta T_R)$ is given by

$$I = A\frac{e}{h}\int \Gamma(E,V)[f_L(E) - f_R(E)]dE, \quad \text{(M12)}$$

Where $h$ is Planck's constant, $f_{(L/R)}(E) = 1/(1 + \exp[(E - \mu_{(L/R)})/k_B T_{(L/R)}])$ is the Fermi-Dirac distribution functions at energy $E$ of the left and right contact respectively, and $0 \leq \Gamma(E, V) \leq 1$ is the single transport mode transmission probability of a ramp-shaped barrier, defined as in Supplementary Fig. 4. The dimensionless scaling constant $A$ accounts for effects present in the experimental data but not in the model, such as multiple transport modes, back gate voltage tuning, finite measurement circuit impedance, etc. It is assumed that $T_L = T_0 + \Delta T$ and $T_R = T_0$ when heating on the left side, and vice versa when heating on the right side.

The transmission probability depends on the barrier length $l$, height $U_{top}$, and effective carrier mass $m^*$, taken to be the one of InAs throughout the barrier region ($m^*$ = 0.0023$m_e$, with $m_e$ the free electron mass). For all data presented in this paper, the transmission probability is calculated fully quantum mechanically (Supplementary section IV.A).

*Adaption to experimental data*

In order to fit equation M1 to our experimental data, three unknown parameters are determined: the temperature difference $\Delta T$ between the two reservoirs, the chemical potential at equilibrium $\mu_0$, and the dimensionless scaling constant $A$. A series of possible pairs of $\mu_0$ and $A$ are determined by minimizing the sum of squared errors *SSE* between the model and experimental data for $\Delta V_{H,L} = \Delta V_{H,R}$ =0 (Supplementary Fig. 6). The model overlaps well with the experimental data only when $\mu_0 < U_{top}$, that is, when the transport is in the tunneling regime. It is not possible to find a unique minimum in the *SSE*. The "line" of minima highlighted in Supplementary Fig. 6 indicates that an increased $\mu_0$ can to some extent be numerically compensated by a decrease in *A*. The adaptation of $\Delta T$ is done for every pair of $\mu_0$ and $A$ along the highlighted line.

For each pair of $\mu_0$ and $A$, it is always possible to uniquely determine the $\Delta T$ that minimizes the SSE between experiment and model, for each applied $\Delta V_H$ (Supplementary Fig. 7). By summing up the *SSE* from the minima of each applied $\Delta V_H$, a total error, *SSE*$_{total}$, is determined for each pair of $\mu_0$ and $A$ (Supplementary Fig. 8).

At this point, $\mu_0$ can be chosen as any value in the range 0~130 meV, with a corresponding $A$ (Supplementary Fig. 8). Because the chemical potential at InAs surfaces are generally known to be pinned in the conduction band [19,20], and a significant backgate voltage of $V_{BG}$ = 10.3 V is applied, we reason that a $\mu_0$ in the larger end of the range that gives a good fit (Supplementary Fig. 8) is more likely. The adaptations used in the main text are thus chosen for $\mu_0$ = 100 meV, and correspondingly $A$ = 710. Note that picking a different value for $\mu_0$ doesn't qualitatively change anything in the adaptation, and only slightly quantitatively changes the relation between $\Delta V_H$ and $\Delta T$ (see adaptations for $\mu_0$ = 80 meV and $\mu_0$ = 130 meV in Supplementary Fig. 9).